\documentclass[aps, twocolumn, pre, showpacs]{revtex4-1}
\usepackage{graphicx}
\usepackage{color}
\usepackage{latexsym}

\def\d {{\rm d}}
\renewcommand{\vec}[1]{{\bf #1}}
\renewcommand{\tensor}[1]{{\bf{{#1}}}}

\begin{document}
\bibliographystyle{prsty}

\title{Robustness of avalanche dynamics in sheared amorphous solids as probed by transverse diffusion}
\author{Joyjit Chattoraj$^{(1)}$}
\author{Christiane Caroli$^{(2)}$}
\author{Ana\"el Lema\^{\i}tre$^{(1)}$}
\affiliation{$^{(1)}$ 
Universit\'e Paris Est -- Laboratoire Navier, ENPC-ParisTech, LCPC, CNRS UMR 8205
2 all\'ee Kepler, 77420 Champs-sur-Marne, France}
\affiliation{$^{(2)}$ INSP, Universit\'e Pierre et Marie Curie-Paris 6, 
CNRS, UMR 7588, 4 place Jussieu, 75252 Paris Cedex 05, France}
\date{\today}

\begin{abstract}
Using numerical simulations, we perform an extensive finite-size analysis of the transverse diffusion coefficient in a sheared 2D amorphous solid, over a broad range of strain rates, at temperatures up to the supercooled liquid regime. We thus obtain direct qualitative evidence for the persistence of correlations between elementary plastic events up to the vicinity of the glass transition temperature $T_g$. A quantitative analysis of the data, combined with a previous study of the $T$- and $\dot\gamma$-dependence of the macroscopic stress~\cite{ChattorajCaroliLemaitre2010}, leads us to conclude that the average avalanche size remains essentially unaffected by temperature up to $T\sim0.75 T_g$.
\end{abstract}
%\pacs{xxx,xxx,xxx}

\maketitle

\section{Introduction}

It is now agreed that, as initially proposed by Argon~\cite{Argon1979}, the macroscopic plastic deformation of amorphous solids is the net result of an accumulation of elementary events which are local rearrangements (``shear transformations'' or ``flips'') of small clusters (``zones'') of atoms, molecules, or particles. Such a flip should be viewed as an Eshelby transformation: since the core cluster is embedded in an elastic medium, its transformation produces a long-ranged elastic field with quadrupolar symmetry. This has been directly observed in numerical simulations~\cite{MaloneyLemaitre2006} and experiments~\cite{SchallWeitzSpaepen2007}.

A flip occurs when a zone (of size $a$) reaches instability at some (local) strain threshold~\cite{MalandroLacks1999,MaloneyLemaitre2004b}. It then starts rearranging into a new stable configuration of lower energy; as described in~\cite{LemaitreCaroli2009} the released energy is evacuated by acoustic radiation into the embedding medium, so that (i) the duration of the event is of order $\tau=a/c_s$, with $c_s$ the shear wave speed; (ii) at a distant point $\vec r$ away from a source at the origin, the long-ranged elastic Eshelby field~\cite{Eshelby1957} is established after the acoustic delay $r/c_s$. This perturbation of the strain field may trigger secondary events, hence may lead to flip-flip correlations and avalanche behavior. As plastic flow progresses, the strain in a given region (zone) in the system is therefore (i) advected by external loading at the imposed strain rate $\dot\gamma$; (ii) subjected to a set of shifts due to elastic signals sent by ongoing flips occurring at random locations in the rest of the system. These elastic signals constitute a self-generated \emph{dynamical noise}, which carries information about flips and controls the nature of the dynamics.

Avalanches were initially evidenced in athermal quasi-static (AQS) simulations. In this limit, thermal fluctuations vanish and the drive is infinitely slow compared to the duration of plastic events, which hence show up as discontinuous drops on the stress-strain curve. On average, each flip releases a macroscopic stress $\mu\,a^d\Delta\epsilon_0/L^d$, with $d$ the space dimension, $L$ the linear size of the system, $\mu$ the shear modulus, and $\Delta\epsilon_0$ a typical strain scale. The amplitude of stress drops in AQS simulations is thus a measure of the avalanche size and was found to be strongly system size-dependent~\cite{MaloneyLemaitre2004a,BaileySchiotzLemaitreJacobsen2007,LernerProcaccia2009}. This permits to conclude that under AQS conditions, flips correlate into avalanches.

But the question then is: to which extent do these correlations survive at finite strain-rates ($\dot\gamma$) and temperatures ($T$)? Indeed, as soon as $\dot\gamma$ is finite, the unfolding of an irreversible event of duration $\tau^{\rm pl}$ spreads over a finite strain interval $\dot\gamma\tau^{\rm pl}$. Hence, plastic events can no longer be identified as discontinuities on the stress-strain curve and, more generally, cannot be isolated. Information about flip correlations and avalanches can no longer be accessed directly. 

Numerical simulations performed in the AQS regime, where avalanches have been identified, have also revealed~\cite{LemaitreCaroli2007,MaloneyRobbins2008} a strong (quasi-linear) system-size dependence of the transverse diffusion coefficient. This motivated a further study~\cite{LemaitreCaroli2009}, where two of us showed that this observable could be used to characterize the correlations between relaxation events. Namely, we measured the transverse diffusion coefficient $D$ in a 2D system driven at finite strain rate in athermal conditions. We then proposed a tentative model leading to a prediction of the $\dot\gamma$ dependence of the average avalanche size and to a scaling expression for $D(\dot\gamma,L)$. Comparing this prediction with extensive simulations data obtained over a broad range of strain rates and systems sizes, we concluded to the validity of the model's predictions, namely to the existence, at finite strain rates, of avalanches of average size $\ell(\dot\gamma)\sim\dot\gamma^{-1/2}$. 

Obviously, the next question to be addressed is that of the effect on avalanche dynamics of a finite temperature~\cite{HentschelKarmakarLernerProcaccia2010}. It has been shown that each flip corresponds to the crossing of a saddle-node bifurcation~\cite{MalandroLacks1999,MaloneyLemaitre2004b}, which occurs after a zone has gradually softened under increasing external loading~\cite{LemaitreCaroli2007}. Thus, near instability, the potential energy landscape (PEL) presents, along one direction corresponding to the shear transformation pathway, a small, gradually decreasing barrier, which vanishes at threshold. When thermal activation is at work, a flip can thus occur ``prematurely'', i.e. before the zone reaches mechanical instability. In reference~\cite{ChattorajCaroliLemaitre2010}, a detailed analysis of the competition between loading and thermal activation led us to propose that, at low temperature, the effect of thermal noise amounts to a rigid downward shift of instability thresholds, while the avalanche dynamics remains unchanged. This yields a prediction for the macroscopic stress $\sigma(\dot\gamma,T)$ which fits quite nicely numerical results over a broad parameter range, thus bringing indirect evidence for the robustness of avalanche dynamics up to a sizeable fraction of the glass transition temperature $T_g$.

In the present paper, we bring further evidence for this conclusion on the basis of a finite-size analysis of transverse diffusion in a 2-dimensional system sheared at finite temperatures and strain-rates.

We first expound, in the following section, the method already used in~\cite{LemaitreCaroli2009} to relate the transverse diffusion coefficient to correlations between plastic events. Our numerical results, reported in Section~\ref{sec:results}, are discussed and interpreted in Section~\ref{sec:discussion}.

\section{Self-diffusion as a probe of flip-flip correlations}
\label{sec:diffusion}

In the following we specialize to the case of two-dimensional systems, and assume for simplicity that:
(i) all flips are identical and characterized by a unique zone size $a$ and a typical scale of strain release $\Delta\epsilon_0$;
(ii) the elastic field associated with any rearrangement can be estimated as the solution of the Eshelby problem~\cite{Eshelby1957} in a homogeneous and isotropic elastic continuum.

We consider the case of a $L\times L$ periodic (Lees-Edwards) system submitted to simple shear at an imposed strain rate $\dot\gamma$. The flow is aligned with the $x$ direction. In view of computing a diffusion coefficient resulting from the plastic activity, we focus on steady state. The condition of stationarity imposes that the total plastic strain release compensates on average the elastic strain increase due to external loading: when the system is strained by $\Delta\gamma=\dot\gamma\Delta t$, since each flip releases a macroscopic strain $a^d\Delta\epsilon_0/L^d$, the average number of flips occurring in a volume of size $L^2$ is,
\begin{equation}
N(\Delta\gamma) = \frac{L^2\,\Delta\gamma}{a^2\,\Delta\epsilon_0}
\end{equation}
This relation translates into an average flip rate:
\begin{equation}
\mathcal{R}=\frac{\dot\gamma L^2}{a^2\,\Delta\epsilon_0}
\quad.
\end{equation}

In order to qualify self-diffusion, we must characterize the long-time behavior of the displacement fluctuations due to the accumulation of Eshelby flips.
The transverse displacement of a particle $i$ between times $t$ and $t+\Delta t$ reads:
\begin{equation}
\label{eqn:deltai}
\Delta y_i(t,t+\Delta t) = \sum_{f\in \mathcal{F}(t,t+\Delta t)} u_y^{\rm E}(\vec r_i-\vec r_f)
\end{equation}
where $u_y^{\rm E}$ is the displacement field generated by an Eshelby source.
The sum runs over the set of all flips, occurring at points $\vec r_f$, whose signals are received at point $\vec r_i$ between times $t$ and $t+\Delta t$. We introduce the source density:
\begin{equation}
\phi_t^{t+\Delta t}(\vec r) = \sum_{f\in \mathcal{F}(t,t+\Delta t)} \delta(\vec r -\vec r_f)
\quad.
\end{equation}

With the above definitions, we write:
\begin{eqnarray*}
\left\langle\Delta y_i^2\right\rangle=&\\
\int \d \vec r_f \d\vec r_f'& \left\langle \phi_{t}^{t+\Delta t}(\vec r_f)\phi_{t}^{t+\Delta t}(\vec r_f')\right\rangle u_y^{\rm E}(\vec r_i-\vec r_f)\,u_y^{\rm E}(\vec r_i-\vec r_f')
\end{eqnarray*}
Thanks to the Lees-Edwards boundary conditions, the system is translationally invariant. In steady flow, the spatial correlation function of the accumulated sources reduces to a function of $\vec r_f-\vec r_f'=\vec R$, and $\Delta t$ only: $\left\langle \phi_{t}^{t+\Delta t}(\vec r_f)\phi_{t}^{t+\Delta t}(\vec r_f')\right\rangle\equiv C(\vec R; \Delta t)$. Whence:
\begin{equation}
\left\langle\Delta y_i^2\right\rangle=
\int\d\vec R \, C(\vec R; \Delta t) \Gamma(\vec R)
\label{eqn:convolution}
\end{equation}
where
\begin{equation}
\Gamma(\vec R)=\int\d \vec r\, u_y^{\rm E}(\vec r)\,u_y^{\rm E}(\vec r-\vec R)
\end{equation}
is the autocorrelation function of the $y$ component of the Eshelby displacement field. 

Athermal simulations on this~\cite{LemaitreCaroli2007,LemaitreCaroli2009} and similar systems~\cite{TanguyLeonforteBarrat2006,MaloneyRobbins2008} have systematically shown convergence towards normal diffusive behavior, ${\left\langle\Delta y_i^2\right\rangle}\propto \Delta t$. We will see that the same holds at finite temperature. In view of equation~(\ref{eqn:convolution}), it implies that at long times, that is for $\Delta t$ much larger that some $\tau^{\rm pl}$ characterizing the temporal decorrelation of plastic activity:
\begin{equation}
\label{eqn:cong}
C(\vec R; \Delta t)\cong \Delta t\,H(\vec R)
\end{equation}
Avalanches are by definition series of correlated flips occurring at distant points. Therefore:
\begin{itemize} 
\item the smallest $\tau^{\rm pl}$ for which the above relation holds is the average avalanche duration
\item the range of $H(\vec R)$ is the average avalanche size $\ell$.
\end{itemize}

The diffusion coefficient $D$ is, finally:
\begin{equation}
D=\lim_{t\to\infty}\frac{\left\langle\Delta y_i^2\right\rangle}{2\,\Delta t}
=\frac{1}{2}\int\d\vec R\,H(\vec R)\,\Gamma(\vec R)
\quad.
\end{equation}
The diffusion coefficient is thus determined by flip-flip correlations, but in some intricate way, which does not grant direct access to $H$ and $\tau^{\rm pl}$. To make further progress, we must therefore introduce assumptions about how correlated sources are organized in space. In the following we consider two situations: (i) completely independent zone flips; (ii) linear avalanches of identical spatial extent $\ell$, composed of flips of uniform density.

\subsection{Independent flips}
\label{sec:eshelby:indep}
If flips are independent, from~(\ref{eqn:deltai}) we can directly write:
\begin{equation}
\label{eqn:diffusion:independent}
\left\langle\Delta y_i^2\right\rangle(\Delta t)=N(\dot\gamma\Delta t)\,
\overline{\left(u_y^{\rm E}\right)^2}
\end{equation}
with the space average: $\overline{\left(u_y^{\rm E}\right)^2}=\frac{1}{L^2}\,\int\d\vec r\,\left(u_y^{\rm E}\right)^2$.
In this case, the sources are delta correlated, whence (with the help of~(\ref{eqn:convolution}),~(\ref{eqn:cong}), and~(\ref{eqn:diffusion:independent})):
\begin{equation}
\frac{C(\vec R; \Delta t)}{\Delta t}=H(\vec R)=\frac{\mathcal{R}}{L^2}\,\delta(\vec R)
\end{equation}

As proposed by~\cite{PicardAjdariLequeuxBocquet2004}, we compute the Eshelby fields as the far-field response to four point-like forces such as depicted on Figure~\ref{fig:forces}. The derivation of relevant formulas is detailed in Appendix A.
\begin{figure}[h]
\includegraphics*[width=0.2\textwidth]{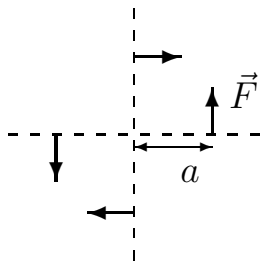}
\caption{The perturbation due to a localized plastic event corresponds to the elastic response
to two force dipoles, of strength $a F =4\mu\Delta\epsilon_0$ (see Appendix).}
\label{fig:forces}
\end{figure}

The displacement field in an infinite medium produced by a source at the origin is:
\begin{equation}
\vec u^{\rm E} = \frac{a^2 \Delta\epsilon_0}{\pi}\,\frac{xy}{r^4}\,\vec r
\quad (r\gg a)
\end{equation}
For $L\gg a$ the transverse displacement fluctuation due to a single flip can be computed at leading order as:
\begin{equation}
\label{eqn:fluctuation}
\overline{\left(u_y^{\rm E}\right)^2}=\frac{a^4\Delta\epsilon_0^2}{8\pi\,L^2}\,\ln(L/a)
\end{equation}
From~(\ref{eqn:diffusion:independent}) and~(\ref{eqn:fluctuation}) we then obtain for the diffusion coefficient:
\begin{equation}
\label{eqn:d:log}
D=\dot\gamma\frac{a^2\Delta\epsilon_0}{16\pi}\,\ln(L/a)
\quad.
\end{equation}

\subsection{Linear avalanches}

Numerical observations in 2D systems provide various pieces of information about avalanche topology, via maps of either relative displacements~\cite{MaloneyLemaitre2004a} or the vorticity field~\cite{MaloneyRobbins2008} in AQS conditions, or the shear strain field in systems sheared at finite strain rate, both at 0~\cite{LemaitreCaroli2009} and finite~\cite{ChattorajCaroliLemaitre2010} temperature. They concur to indicate that avalanches form quasi-linear patterns, oriented close to the $x$ and $y$ axes in the simple shear geometry, i.e. at $\pi/4$ of the principal axes of the strain tensor. This is consistent with the quadrupolar structure of the Eshelby strain field, which reads, in polar coordinates:
\begin{equation}
\epsilon_{xy} = \frac{a^2\Delta\epsilon_0}{\pi}\,\frac{\cos(4\theta)}{r^2}
\end{equation}
That is, the strain shift following a flip at the origin is maximum for target zones located along the $x$ and $y$ axes.

This motivates us to model avalanches as linear structures of identical extent $\ell$, composed of flips of uniform linear density $\nu$, and aligned with equal probabilities along the $x$ and $y$ axes. Since each avalanche involves $n=\nu\ell$ flips, the average number of avalanches occurring over an strain interval $\Delta\gamma$ is:
\begin{equation}
\label{eqn:na}
N_a(\Delta\gamma)= \frac{L^2\,\Delta\gamma}{\nu\ell\,a^2\,\Delta\epsilon_0}
\end{equation}
while the avalanche rate is:
\begin{equation}
\mathcal{R}_a(\Delta\gamma)= \frac{L^2\,\dot\gamma}{\nu\ell\,a^2\,\Delta\epsilon_0}
\quad.
\end{equation}
The particle displacement (see Eq.~(\ref{eqn:deltai})) can now be rewritten as a sum over independent avalanches, leading to:
\begin{equation}
\label{eqn:diffusion:avalanche}
\left\langle\Delta y_i^2\right\rangle(\Delta t)=N_a(\dot\gamma\Delta t)\,
\frac{1}{2}\left(\overline{\left(u_y^{\rm A,x}\right)^2}
+\overline{\left(u_y^{\rm A,y}\right)^2}\right)
\end{equation}
where e.g. $u_y^{\rm A,x}$ is the sum of the Eshelby fields of the flips composing an avalanche along O$x$:
\begin{equation}
\overline{\left(u_y^{\rm A,x}\right)^2}=\frac{\nu^2}{L^2}\,
\int_{-\ell/2}^{\ell/2}\int_{-\ell/2}^{\ell/2}\d x \d x'\,\Gamma\left[(x-x')\,\vec e_x\right]
\end{equation}
with $\vec e_x$ the unit vector in the $x$ direction.
A similar expression holds for $y$-avalanches.

We show in the Appendix that:
\begin{equation}
\Gamma(\vec R=(R,\theta)) = \frac{a^4\Delta\epsilon_0^2}{16\pi}\,\int_{R/L}^{\infty} \frac{\d z}{z}\,G(z,\theta)
\end{equation}
where,
\begin{eqnarray*}
G(z,\theta)&=& 2\,J_0(z)
-3\,\cos(2\theta)\,J_2(z)\\
&&+2\,\cos(4\theta)\,J_4(z)
-\cos(6\theta)\,J_6(z)
\end{eqnarray*}
with $J_n$ the Bessel functions.
To lowest order in $\ell/L$ this yields:
\begin{equation}
\frac{1}{2}\left(\overline{\left(u_y^{\rm A,x}\right)^2}
+\overline{\left(u_y^{\rm A,y}\right)^2}\right)=\frac{a^4\Delta\epsilon_0}{8\pi}\,\frac{\ell^2}{L^2}\,\ln(L/\ell)
%\overline{\left(u_y^{\rm A,y}\right)^2}=\frac{\nu^2}{L^2}\,
\end{equation}

Finally, using~(\ref{eqn:na}) and~(\ref{eqn:diffusion:avalanche}) we find:
\begin{equation}
\label{eqn:d}
D=\dot\gamma\frac{a^2\Delta\epsilon_0}{16\pi}\,\nu \ell\,\ln(L/\ell)
\quad.
\end{equation}

\section{Numerical results}
\label{sec:results}
We report simulation results obtained on a 2D Lennard-Jones system composed of small (S) and large (L) particles with equal masses $m=1$, radii $R_L=0.5$, $R_S=0.3$ (we work in standard LJ units), and number ratio $N_L/N_S=(1+\sqrt{5})/4$. The packing fraction of our $L\times L$ system is $\pi(N_LR_L^2+N_SR_S^2)/L^2=0.9$.

Finite temperature simulations are performed using velocity rescaling.
To characterize the relaxation behavior of this system, we have obtained, for system size $L=40$, equilibrium states, by progressively lowering the temperature starting from the liquid state at $T=1$. No crystallization occurs. We measure a nominal glass transition temperature as that where the time $\tau_\alpha$ (defined from the relaxation of the incoherent scattering function of large particles) reaches $10^4$. With this criterion, $T_g\cong0.28$.

The system is submitted to simple shear using Lees-Edwards boundary conditions, at imposed strain rates ranging from $\dot\gamma=10^{-5}$ to $10^{-2}$. We present measurements of the transverse diffusion coefficient for various temperatures ranging from $T=0.05$ up into the supercooled liquid regime. All the data presented here are obtained after 100\% preshearing to ensure that our systems are in steady state.

Finite size analysis is performed using systems of linear sizes $L=10,20,40,80,160$. Statistical accuracy turns out to demand large sets and long strain intervals: for example, twenty-five $L=40$ (2837 particles) systems have been strained up to 1300\% and five $L=160$ (45395 particles) systems up to 2400\%. This entails a heavy numerical cost: at our lowest $\dot\gamma=10^{-5}$, and for each value of $T$, straining our five $L=160$ systems by 2400\% using $dt=0.01$ requires a total $1.2\times10^9$ time-steps; with $2.5\mu$s/particle/time-step on recent clusters, this amounts to $\sim40000$ hours.

We will find convenient in the following to introduce a \emph{reduced diffusion coefficient}: 
\begin{equation}
\widehat D=D/\dot\gamma
\end{equation}
which measures the growth of fluctuations with strain (instead of time).

\begin{figure}[h]
\includegraphics*[width=0.45\textwidth]{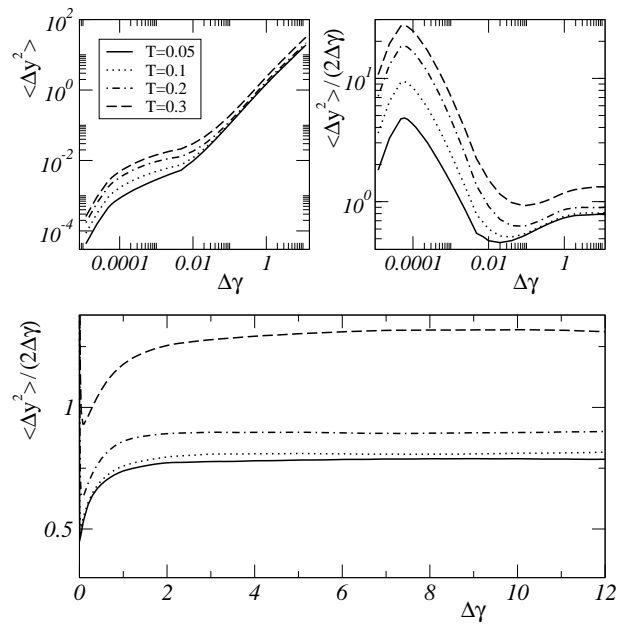}
\caption{Transverse displacement fluctuation for $L=40$, $\dot\gamma=4\times 10^{-4}$, $T=0.05$, 0.1, 0.2, and 0.3.
}
\label{fig:2}
\end{figure}

We present on Fig.~\ref{fig:2} transverse diffusion data for system size $L=40$, for a single value of the strain rate $\dot\gamma=4\times10^{-4}$, and for temperatures $T=0.05$, 0.1, 0.2 and 0.3.
Panel~\ref{fig:2}(a) shows a log-log plot of $\langle \Delta y^2\rangle$ vs $\Delta\gamma$. These curves exhibit the three usual regimes: a quadratic behavior at very short times (strains) corresponding to the initial thermal exploration of the cage; a caging phase showing up as a quasi-plateau; finally, normal diffusive behavior as particles start escaping from their cages. The same data are replotted as $\langle \Delta y^2\rangle/(2\Delta\gamma)$ vs $\Delta\gamma$ on both panels~\ref{fig:2}(b) and~\ref{fig:2}(c).
The log-log plot (\ref{fig:2}(b)) provides details about the transient behavior, the caging phase corresponding to the decrease following the initial peak. The lin-lin plot (\ref{fig:2}(c)) emphasizes the late, normal, diffusive behavior reached, as in the athermal case~\cite{LemaitreCaroli2009}, after a transient of extent $\Delta\gamma\sim 1$. The amplitude of the asymptotic plateau defines the reduced diffusion coefficient $\widehat D$. 

\begin{figure}[h]
\includegraphics*[width=0.45\textwidth]{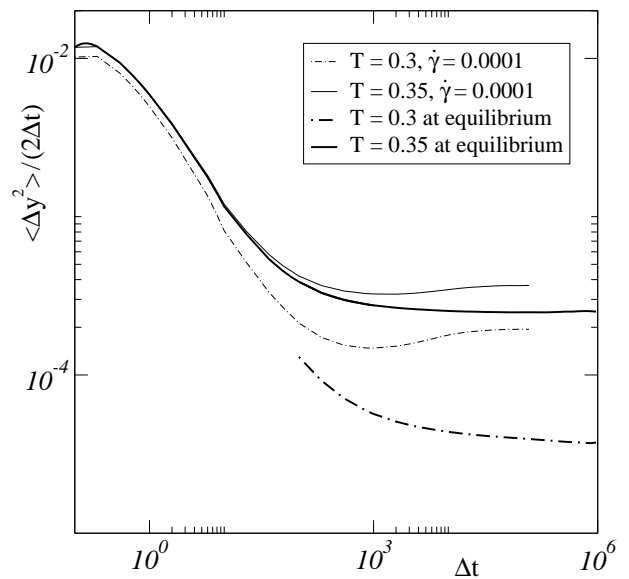}
\caption{Comparison between transverse displacement fluctuations in the sheared ($\dot\gamma=10^{-4}$) vs equilibrated, unsheared, system, for size $L=40$, $T=0.3$ and 0.35.}
\label{fig:3}
\end{figure}

Note that the highest temperature investigated in these graphs, $T=0.3$, lies closely above our measured $T_g\cong0.28$, i.e. belongs to the supercooled liquid regime. At this temperature, we can also measure diffusion in the equilibrated unsheared system. On figure~\ref{fig:3}, we plot $\langle \Delta y^2\rangle/(2\Delta t)$ vs $\Delta t$ for both the sheared and unsheared systems at two values of temperature, $T=0.3$ and $0.35$. One clearly sees that, up to the end of the caging regime, transverse motion is only very weakly affected by shearing. By contrast, at later times, diffusion is enhanced by shear: this effect is already sizeable ($\sim50\%$) at $T=0.35$, and becomes more conspicuous as temperature decreases towards $T_g$. It is also visible on this figure that, under shear, the crossover between the caging and diffusive regimes, as signalled by the minimum of $\langle \Delta y^2\rangle/(2\Delta t)$, is barely sensitive to the increase of the alpha relaxation time ($\tau_\alpha\sim 120$ for $T=0.35$ and $\sim2000$ for $T=0.3$).

\begin{figure}[h]
\includegraphics*[width=0.45\textwidth]{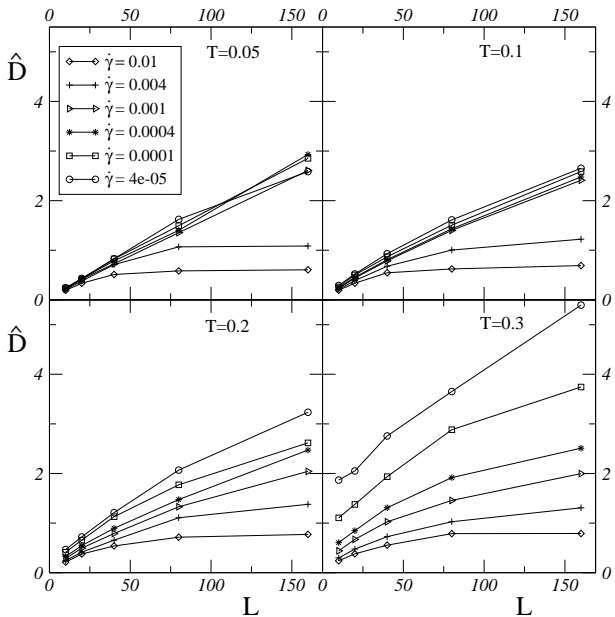}
\caption{Reduced diffusion coefficient $\widehat D$ vs system size $L$, for different strain rates and temperatures.}
\label{fig:4}
\end{figure}

The complete set of our diffusion data as a function of system size, temperature, and strain rate, is displayed on figure~\ref{fig:4}. Each panel corresponds to a different, fixed, temperature, and shows a plot of $\widehat D$ vs system size, for $\dot\gamma$ ranging from $4\times10^{-5}$ to $10^{-2}$. As is immediately seen, $\widehat D$ is noticeably size-dependent up to the highest temperature, $T=0.3$, which lies within the supercooled regime. 

At $T=0.05$, we recover the features previously observed in our athermal simulations (see~\cite{LemaitreCaroli2009} and discussion below), namely: (i) at fixed $L$, $\widehat D$ becomes nearly $\dot\gamma$-independent for the lower $\dot\gamma$'s; (ii) at these low strain rates, the size dependence of $\widehat D$ is quasi-linear; and (iii) it becomes much weaker, quasi-logarithmic, at the higher $\dot\gamma$'s. Increasing temperature up to 0.1 does not significantly alter these functional forms, nor does it induce a noticeable change in the magnitude of $\widehat D$.

Thermal effects become clearly visible at $T=0.2$, that is rather close below $T_g$, where we observe a splay of the $\widehat D(L)$ curves at all $\dot\gamma$'s. This is accompanied by a change in the $L$-dependence of $\widehat D$, which becomes sub-linear in the whole $\dot\gamma$ range that we can access. These effects become even more conspicuous upon crossing the glass transition as seen on panel~(d).

\begin{figure}[h]
\includegraphics*[width=0.49\textwidth]{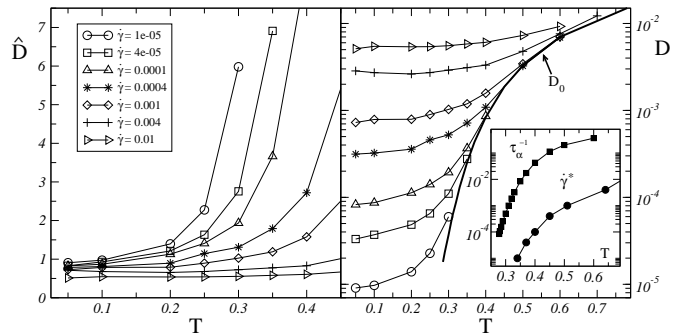}
\caption{Diffusion coefficient vs temperature for different strain rates. (a): reduced coefficient $\hat D$; (b): $D$ compared with $D_0$ as measured in the equilibrated supercooled liquid; insert: temperature dependence of $\dot\gamma^\star$ (see text) compared with that of $\tau_\alpha^{-1}$.}
\label{fig:dhatandd}
\end{figure}

A plot of $\widehat D$ versus temperature is presented on Figure~\ref{fig:dhatandd}(a), for $L=40$, and for the different $\dot\gamma$'s used in this study. Clearly, $\widehat D$ increases with temperature, which is expected as thermal fluctuations increasingly contribute to diffusion. This increase is hardly visible at the highest strain rates and becomes prominent at low $\dot\gamma$: this is largely due to the definition of $\widehat D=D/\dot\gamma$. Indeed, the reduced coefficient $\widehat D$ is the appropriate measure of diffusion in the low-$T$, low-$\dot\gamma$ limit, where strain controls particle motion, while $D$, the standard diffusion coefficient, better characterizes diffusion at higher temperatures, in the supercooled regime, where the unsheared system presents normal diffusive behavior. Near the glass transition, $\widehat D$ therefore captures the contribution of thermal fluctuations as accumulated over time intervals $\propto\dot\gamma^{-1}$, whence its enhanced $T$-sensitivity at the lowest $\dot\gamma$'s.

The cross-over between a strain-controlled and a temperature-controlled diffusive regimes is illustrated on Fig.~\ref{fig:dhatandd}(b) by a plot of the same data set as $D(T)$ for different $\dot\gamma$'s. On the same graph we also plot values of the diffusion coefficient (denoted $D_0$) measured in the equilibrated, unsheared, system down to $T=0.278$. For each $\dot\gamma$, $D(T;\dot\gamma)$ increases with $T$ and merges at high temperatures with the equilibrium curve $D_0(T)$. The merging points $\dot\gamma^\star(T)$ are plotted in the inset of Fig.~\ref{fig:dhatandd}(b): they define a cross-over line delimiting a low strain-rate, high-temperature, region where thermal fluctuations effects largely dominate those of mechanical noise. Onuki and Yamamoto~\cite{OnukiYamamoto1998,YamamotoOnuki1998a} suggested that the dynamics of the sheared system should merge with that of the equilibrated supercooled liquid at a cross-over defined by $\dot\gamma^\star\sim \tau_\alpha^{-1}$, with $\tau_\alpha$ the $\alpha$-relaxation time. We thus also report in the inset of Fig.~\ref{fig:dhatandd}(b) our measured values of $\tau_\alpha^{-1}(T)$. Both $\dot\gamma^\star$ and $\tau_\alpha^{-1}$ strongly increase with $T$ and are roughly parallel on the log-log plot, though shifted by more than 2 decades. Namely, the cross-over criterion corresponds to $\dot\gamma^\star\tau_\alpha\sim10^{-2}$--$10^{-3}$. This result is consistent with the findings of Furukawa~{\it et al}~\cite{FurukawaKimSaitoTanaka2009}, who studied the cross-over between Newtonian and non-Newtonian rheological regimes on a similar 2D LJ system.

\section{Discussion}
\label{sec:discussion}

\begin{figure}[h]
\includegraphics*[width=0.45\textwidth]{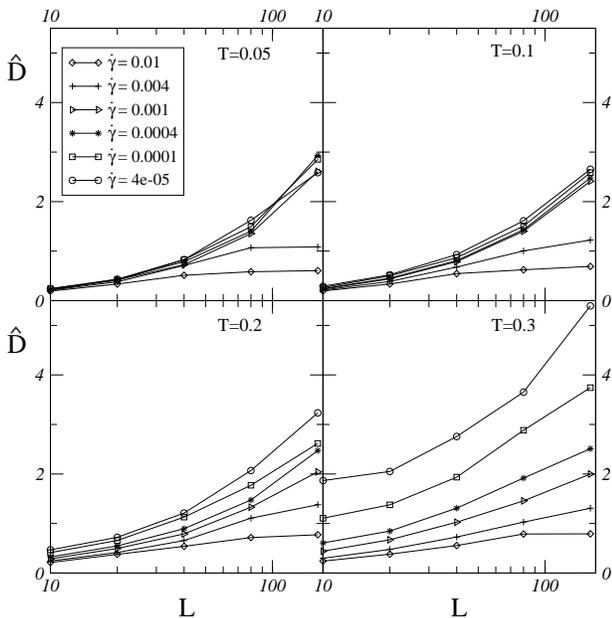}
\caption{Log-lin plots of the same data as on Fig.~\ref{fig:4}.}
\label{fig:8}
\end{figure}

Our aim here is to use diffusion data in order to probe the existence of correlations between plastic events at finite temperatures. We have shown in Section~\ref{sec:eshelby:indep} that the complete absence of correlations directly translates into the $\hat D\propto \log L$ behavior. Hence, any departure from this scaling indubitably signals the presence of flip-flip correlations. We thus present on figure~\ref{fig:8} plots of $\widehat D$ vs $\log L$. Strikingly, for the four temperatures considered, the $\widehat D\propto\log L$ behavior is only found, as in the athermal limit, at the highest $\dot\gamma$'s. At lower strain rates, the growth of $\widehat D$ is clearly faster than a logarithm. This direct proof for the persistence of some degree of correlation, up to $T=0.2\approx 0.75T_g$ and even $T=0.3$ (which lies above the nominal glass transition temperature) is the primary outcome of this work.

In a recent article~\cite{ChattorajCaroliLemaitre2010}, we analyzed the $T$ and $\dot\gamma$-dependence of the macroscopic stress for the same system. We proposed a model in which, at low temperatures, the effect of thermal fluctuations reduces to a mere lowering of the strains at which plastic events occur, while the avalanche size remains essentially identical to that in the $T=0$ limit. We found that numerical data could be matched with the prediction of this model for the macroscopic shear stress, $\sigma(\dot\gamma,T)$, over a broad range of temperatures extending up to $T\simeq0.2$. This indirect argument led us to conclude that the avalanche dynamics is essentially unchanged up to this temperature.
 
Here, we bring direct, yet qualitative, evidence for the persistence of correlations between plastic events up to (and even beyond) the glass transition. The question then arises whether diffusion measurements permit to draw more quantitative conclusions about the effect of temperature on the scale of spatial correlations, i.e. on the average avalanche size. In particular, is it possible to provide a further test of whether $\ell(\dot\gamma)$ is essentially $T$-independent up to $T=0.2$?

\begin{figure}[h]
\includegraphics*[width=0.45\textwidth]{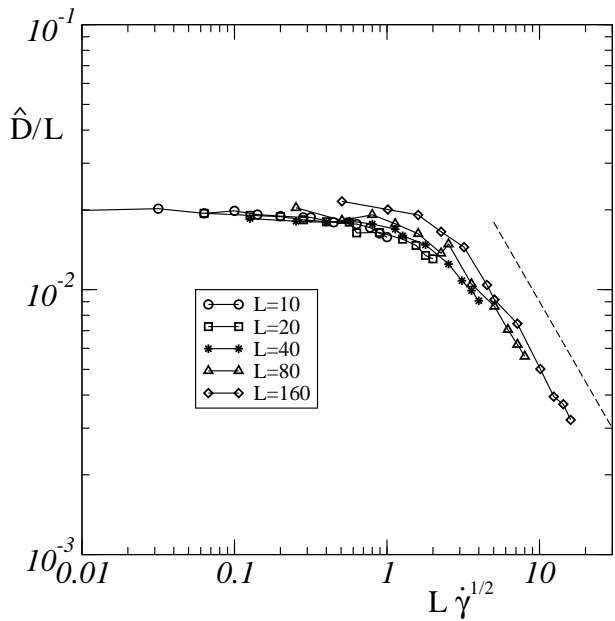}
\caption{Scaling plot of transverse diffusion data from athermal simulations on the same system. The dashed line has slope $-1$.}
\label{fig:athermal}
\end{figure}

Two of us have previously proposed a model for the $\dot\gamma$-dependence of the avalanche size in the $T=0$ limit~\cite{LemaitreCaroli2009}, based on an analysis of the mechanical noise generated by the flips themselves. This model considers a given zone, and separates the noise it receives into: (i) signals originating from nearby flips (within a region of radius $R$); (ii) background noise, emanating from all other, more distant, flips. The avalanche length $\ell$ is defined as the largest $R$ such that nearby signals are able to bias the occurrence of secondary events, i.e. to give rise to correlations between flips. This demands that nearby signals (i) do not overlap, (ii) stand out of the background noise accumulated during a flip duration. Both conditions yield the common prediction, $\ell\sim\dot\gamma^{-1/2}$, down to a size-dependent cross-over strain rate $\dot\gamma_c(L)\sim1/L^2$, below which it saturates to $\ell\sim L$. From Eq.~(\ref{eqn:d}), it entails that the transverse diffusion data for systems of various sizes should obey the scaling relation $D/L=f(L\,\sqrt{\dot\gamma})$, a prediction which was found in~\cite{LemaitreCaroli2009} to be very well matched by numerical data. The corresponding scaling plot, shown on Fig.~\ref{fig:athermal}, illustrates the quality of the data collapse in the athermal limit. Above the cross-over, the master curve $f(L\sqrt{\dot\gamma})$ exhibits the $f(x)\sim 1/x$ predicted from the above argument. The large plateau at low $x$ corresponds to the regime where the avalanche length reaches the system size.

\begin{figure}[h]
\includegraphics*[width=0.45\textwidth]{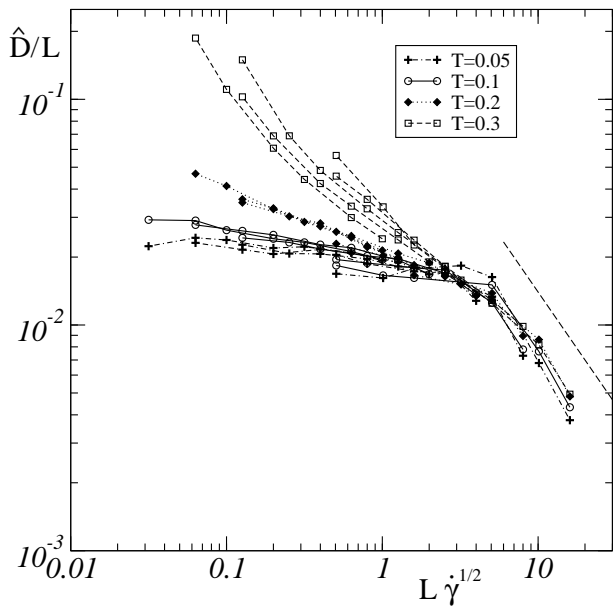}
\caption{Scaling plot of transverse diffusion data for temperatures $T=0.05$, 0.1, 0.2, and 0.3. Each curve corresponds to a single system size ($L=10$, 20, 40, 80, 160). Dashed line: slope -1.}
\label{fig:collapse}
\end{figure}

To evaluate the importance of thermal effects on $\ell$, we now attempt the same kind of collapse using our finite $T$ data. The results presented on Fig.~\ref{fig:collapse} clearly show a cross-over around $x_c=L\,\sqrt{\dot\gamma_c(L)}\sim2$--$3$. Above $x_c$ the whole set of data collapse onto a single curve, with slope $-1$ in the log-log plot. This is precisely the scaling behavior obeyed by athermal data: it is the signature of a regime where the avalanche size scales as $1/\sqrt{\dot\gamma}$. The collapse found here shows that, in this regime, the avalanche size, hence the avalanche dynamics, is roughly unaffected by temperature. This is consistent with our observation, illustrated on Fig.~\ref{fig:dhatandd}, that for each temperature, it is at the higher $\dot\gamma$'s that mechanical noise dominates thermal noise. 

Below cross-over, a very different behavior emerges. For the three temperatures $T=0.05$, 0.1, 0.2, each set of data continues to collapse, but now onto a different master curve $f_T(x)$ for each value of $T$. In contrast, there is a clear splay of the $T=0.3$ data for different system sizes. 

At the lowest temperature $T=0.05$, the overall master curve exhibits exactly the same behavior as found in the athermal limit: up to this temperature, of order $T_g/5$, and down to the lowest $\dot\gamma$'s that we can investigate, temperature has a negligible effect. In particular, the quasi-plateau seen below the cross-over, corresponding to the $\widehat D\propto L$ behavior (see low-strain-rate data on Fig.~\ref{fig:4}(a)), indicates that, in this regime, the avalanche length saturates at $\ell\sim L$.

As $T$ increases to 0.1, then to 0.2, below cross-over, $f_T$ remains quasi-linear in the log-log plot, but develops an increasing negative slope. Indeed, as expected, the effect of thermal fluctuations on the \emph{reduced} diffusion coefficient $\widehat D$ increases with decreasing $\dot\gamma$. Yet diffusion data alone do not allow us to conclude whether or not these variations of $\widehat D$ with $T$ correspond to alterations of the avalanche size. However, as mentioned above, another piece of information is available to us: we have shown that up to $T=0.2$, the macroscopic stress data could be fitted by a model in which the avalanche dynamics is essentially unaffected by temperature~\cite{ChattorajCaroliLemaitre2010}. We are therefore led to attribute the growth of $\widehat D(\dot\gamma)$ with $T$ below cross-over to thermally activated processes which do not contribute, on average, to the relaxation of the shear stress. 

In our view, each plastic event strongly reshuffles atoms in its close vicinity, thus bringing the system into a new region of the PEL, with a finite density of small energy barriers~\cite{UtzDebenedettiStillinger2000}. As thermal relaxation from these ``rejuvenated'' configurations is constantly reinitialized by steady plastic deformation, it yields a finite contribution to particle diffusion. Our observation that macroscopic stress is insensitive to the occurrence of such extra relaxation events entails that these events are enslaved to the plastic dynamics, i.e. do not feed-back into it. It also indicates that the small-amplitude ruggedness of the PEL is essentially unbiased by the macroscopic stress.

At temperatures such that the diffusion coefficient is non-vanishing in the absence of stress, i.e. when thermal noise allows the system to fully explore its PEL, thermally activated events are likely to feed back into the flips' dynamics itself. This is when a splay develops on the scaling plot (see $T=0.3$ data on Fig.~\ref{fig:collapse}).\\

In summary, the primary outcome of this study is that in a two-dimensional LJ sheared system at finite temperatures and strain rates, correlations between elementary plastic events do persist up to the vicinity of the glass transition: they show up as a stronger-than-log size dependence of the diffusion coefficient. Moreover, the collapse of rescaled diffusion data above the cross-over $\dot\gamma_c\sim1/L^2$ (see Fig.~\ref{fig:collapse}), valid up to the glass transition, leads us to conclude that avalanches are unaffected by temperature in the shear-controlled regime $\dot\gamma>\dot\gamma^\star(T)$ (see insert of Fig.~\ref{fig:dhatandd}(b)). This brings further support to our previous conclusion, based on the analysis of macroscopic rheology, that up to $T\sim0.75 T_g$ the average avalanche size remains essentially unaffected by temperature. Indeed, for all the strain rates which have been studied, and for $T\le 0.75 T_g$, the system is in the shear-controlled diffusive regime. It is only upon approaching $T_g$, and for $\dot\gamma\,\tau_\alpha\lesssim10^{-2}$--$10^{-3}$, that the effects of mechanical and thermal noise can no longer be unraveled. It is then likely that, in this regime, thermal noise gradually destroys correlations. 

Finally, the thermal enhancement of diffusion found at the lower strain rates (below cross-over on Fig.~\ref{fig:collapse}), can reasonably be assigned to rejuvenation of local configurations due to the plastic activity, which permanently feeds additional thermal relaxation. 

\thanks

The authors wish to thank the French competitiveness cluster Advancity and Region \^Ile de France for their financial help. 
This work was granted access to the HPC resources of IDRIS under the allocation 2010-99644 made by GENCI (Grand Equipement National de Calcul Intensif).

\appendix
\section{Appendix}

The Eshelby field corresponding to a flip can be viewed, following Picard~{\it et al\/}~\cite{PicardAjdariLequeuxBocquet2004}, as the far-field response to a set of four forces, applied near the origin in an infinite, incompressible, elastic medium, as depicted on figure~\ref{fig:forces}:
\begin{equation}
\vec u^E(\vec r)=\sum_{i=1}^4 \tensor Q(\vec r-\vec r_i).\vec F_i
\end{equation}
where $\tensor Q$ is the Green's tensor corresponding to the response to a point force located at the origin~\cite{LandauLifshitz1975}:
\begin{equation}
\vec Q = \frac{1}{4\pi\mu}\,\left(-\ln r\, \tensor I + \frac{1}{r^2}\,\vec r\,\vec r\right)
\end{equation}
with $\mu$ the shear modulus and $\tensor I$ the unit tensor. An expansion in $a/r$ yields at lowest order:
\begin{equation}
\vec U = \frac{a F}{2\,\pi\mu}\,\frac{xy}{r^4}\vec r
\end{equation}
To relate the dipolar strength $a F$ to a strain release scale, we compare the stress generated by the point forces, $F/2a$, with that, $2\mu\Delta\epsilon_0$, corresponding to the local strain release within the zone. This yields the relation:
\begin{equation}
\vec U = \frac{a^2\Delta\epsilon_0}{\pi}\,\frac{xy}{r^4}\vec r
\end{equation}

The function $\Gamma$ is next evaluated as:
\begin{eqnarray*}
\Gamma(\vec R)&=&L^2\,\overline{u_y^{\rm E}(\vec r)u_y^{\rm E}(\vec r-\vec R)}\\
&=& \frac{1}{(2\,\pi)^2}\,\int \d\vec q\,|\hat u_y^{\rm E}(\vec q)|^2\,e^{-i\vec q.\vec R}
\end{eqnarray*}
Using,
\begin{equation}
\hat u_y^{\rm E}(\vec q)=-{ia^2 \Delta\epsilon_0}\,\frac{q_x\,(q_x^2-q_y^2)}{q^4}
\end{equation}
we find:
\begin{equation}
\Gamma(\vec R)=\frac{a^4\,\Delta\epsilon_0^2}{4\pi^2\,L^2}\,\int_{q_{\rm min}}^\infty \frac{\d q}{q}\,\int_0^{2\,\pi}\,\d\theta'\,
\gamma(\theta')\,e^{-iq R\,\cos(\theta'-\theta)}
\end{equation}
where $\vec R = (R,\theta)$ in polar coordinates, with the lower cut-off $q_{\rm min}\sim 1/L$, and:
\begin{equation}
\gamma(\theta')=\cos^2\theta'\left(\cos^2\theta'-\sin^2\theta'\right)^2
\quad.
\end{equation}
We finally obtain:
\begin{equation}
\Gamma(\vec R) = \frac{a^4\Delta\epsilon_0^2}{16\pi}\,\int_{R/L}^{\infty} \frac{\d z}{z}\,G(z,\theta)
\end{equation}
where,
\begin{eqnarray*}
G(z,\theta)&=& 2\,J_0(z)
-3\,\cos(2\theta)\,J_2(z)\\
&&+2\,\cos(4\theta)\,J_4(z)
-\cos(6\theta)\,J_6(z)
\end{eqnarray*}
with $J_n$ the Bessel functions.

\end{document}